\documentclass[twocolumn,showpacs,preprintnumbers,amsmath,amssymb]{revtex4-1}
\usepackage{amssymb}
\usepackage{txfonts}

\usepackage{graphicx}
\usepackage{dcolumn}
\usepackage{bm}

\begin{document}

\title{Solution and testing of the Abraham-Minkowski controversy of light-atom interacting system}

\author{Wen-Zhuo Zhang, Peng Zhang, Ru-Quan Wang, and Wu-Ming Liu$^*$}
\affiliation{Beijing National Laboratory for Condensed Matter Physics, Institute of Physics, Chinese Academy of Sciences, Beijing 100190, P.R. China}
\email{wmliu@iphy.ac.cn}
\date{\today}

\begin{abstract}
We present the origin of the Abraham-Minkowski controversy of light-matter wave interacting system, which is a special case of the centaury-old Abraham-Minkowski controversy. We solve the controversy of laser-atom interacting case and find that for systems with perfect atomic spatial coherence, the systems prefer to show Minkowski momentum and canonical momentum for the atoms and light, respectively; while for the systems where the atoms are spatially incoherent, the momenta of light and atoms would choose the Abraham and kinetic forms. The provement of our solution can be realized with current techniques, using three-dimensional optical lattices and electromagnetically-induced absorption (EIA) to distinguish the kinetic and canonical recoil momentum of ultra-cold atoms.
\end{abstract}

\maketitle

The Abraham-Minkowski controversy \cite{Leonhardt,RMP,AOP,RSA} is a centaury-old problem which focuses on the momentum of light in optical media. Minkowski states that it is $\int(\mathbf{D}\times\mathbf{B})dV=n\hbar\mathbf{k_0}$ \cite{Min}, while Abraham argues that it should be $\int(\mathbf{E}\times\mathbf{H}/c^2)dV=\hbar\mathbf{k_0}/n$ \cite{Abr}, with $\mathbf{k_0}$ the wave vector of light in vacuum and $n$ the refractive-index of the medium. A recent experiment measuring the recoil momentum of ultra-cold atoms supports Minkowski's view \cite{MIT}, while another experiment measuring the recoil momentum of silica filaments prefers Abraham's one \cite{ZSU}. Recently, Barnett resolved the long-standing problem \cite{Barnett}, and demonstrated that the Abraham momentum is indeed the kinetic momentum of light which equals to $\hbar\mathbf{k_0}/n_g$ where $n_g$ is the group refractive-index of the medium, while the Minkowski momentum is indeed the canonical momentum of the light which equals to $n_p\hbar\mathbf{k_0}$ with $n_p$ the phase refractive-index of the medium.

Barnett's resolution shows that the unique total momentum $\mathbf{P}$ of the light-optical medium interacting system can be expressed in two different combinations \cite{Barnett,Hinds,Leonhardt-pra}
\begin{equation}\label{m}
\mathbf{P}=\mathbf{P_k}+\mathbf{P_{Abr}}=\mathbf{P_c}+\mathbf{P_{Min}},
\end{equation}
where $\mathbf{P_k}$ and $\mathbf{P_c}$ are the kinetic and canonical momenta of the optical medium, respectively; $\mathbf{P_{Abr}}$ and $\mathbf{P_{Min}}$ are the Abraham and Minkowski momenta of the light \cite{Barnett}. Although the form of the Abraham and Minkowski momenta of light is fixed, the reason why the system chooses the combination $\mathbf{P_k}+\mathbf{P_{Abr}}$ in some cases while $\mathbf{P_c}+\mathbf{P_{Min}}$ in other cases is still unknown.

In this paper, we consider the special case in which the optical medium can be described by a \emph{matter wave}. We present the origin of the Abraham-Minkowski controversy of light-matter wave interacting systems, and solve it in the light-atom case for the first time. Our solution answers the most \emph{essential} question that why the light-atom interacting system shows $\mathbf{P_c}+\mathbf{P_{Min}}$ in some cases, while $\mathbf{P_k}+\mathbf{P_{Abr}}$ in other cases. We design an experiment to prove the solution by measuring both the recoil $\mathbf{P_c}$ and $\mathbf{P_k}$ of ultra-cold $^{87}$Rb atoms in a single setup, which is more advanced than previous experiments in which each setup can only measure either recoil $\mathbf{P_c}$ or $\mathbf{P_k}$ of the optical media.

\emph{Origin of the Abraham-Minkowski controversy of light-matter wave interacting system}.
A free matter wave, which could be a charged particle, an atom, or a pure Bose-Einstein condensate (BEC), has only one unique momentum $\mathbf{P_0}=\hbar\mathbf{k}$. There is no difference between its $\mathbf{P_c}$ and the $\mathbf{P_k}$ and thus no Abraham-Minkowski controversy for it. However, when the interaction with light is involved, controversy appears. For a charged particle coupling with an electromagnetic field, its kinetic momentum is $\mathbf{P_k}=\mathbf{P_c}-e\mathbf{A}$, with $\mathbf{P_c}$ is its canonical momentum and $\mathbf{A}$ is the vector potential of the light. The Hamiltonian can be written
\begin{equation}
H=\sqrt{(\mathbf{P_c}-e\mathbf{A})^2c^2+m_0^2c^4}+e\Phi,\label{h}
\end{equation}
where $\Phi$ is the scalar potential of the light.

Eq.~(\ref{h}) shows that the charged particle couples the four-potential $(\mathbf{A},\Phi)$ of the light by momentum part $-e\mathbf{A}$ and energy part $e\Phi$. $\mathbf{P_c}$ acts as the particle's ``free" momentum and $\mathbf{P_k}$ acts as its ``dressed" momentum. Since the total energy of the light-particle interacting system is equal to Eq.~(\ref{h}) plus the ``free" energy of light $\frac{1}{2}(\varepsilon_0\mathbf{E}^2+\mathbf{B}^2/\mu_0)$, the total momentum of the system needs the ``free" momentum of the light, which is just the Abraham momentum $\mathbf{P_{Abr}}$ of the light \cite{Feigel,OC}. Therefore the unique total momentum of the system is
\begin{equation}\label{tm}
\mathbf{P}=\mathbf{P_c}-e\mathbf{A}+\mathbf{P_{Abr}}.
\end{equation}
The Minkowski momentum of light $\mathbf{P_{Min}}$ here is apparently $\mathbf{P_{Abr}}-e\mathbf{A}$. From Eq.~(\ref{tm}) we can see whether to write the unique total momentum $\mathbf{P}$ as $\mathbf{P_k}+\mathbf{P_{Abr}}$ or as $\mathbf{P_c}+\mathbf{P_{Min}}$, depends on whether adding the coupling momentum $-e\mathbf{A}$ to $\mathbf{P_c}$ to form $\mathbf{P_k}$, or to $\mathbf{P_{Abr}}$ to form $\mathbf{P_{Min}}$. This directly induces the Abraham-Minkowski controversy of the light-particle interacting system. It should be emphasized that $\mathbf{P_c}$ of the charged particle and $\mathbf{P_{Abr}}$ of the light are not equal to their unperturbed values in free space. The reason is that when the quantization is made in the light-particle interacting system, the momentum eigenvalues of the charged particle and light are not the same as their unperturbed ones in free space. In quantum theory, $\mathbf{P_c}$ and $\mathbf{P_{Min}}$ are associated with wavelength and considered as the wave-like limit of the system, while $\mathbf{P_k}$ and $\mathbf{P_{Abr}}$ are associated with kinetic energy and considered as the particle-like limit of the system \cite{Leonhardt-pra, Hinds}. Therefore, we seek to study the condition in which the system prefers to show its total momentum as $\mathbf{P_c}+\mathbf{P_{Min}}$ or $\mathbf{P_k}+\mathbf{P_{Abr}}$ to solve the controversy, and find available experimental schemes to test our solution. However, the light-particle interacting system is incapable to do this, because only the unperturbed charged particle and photon status can be detected for the initial and final states of the S-matrix in quantum electrodynamics (QED). Fortunately, for an light-atom interacting system, things are different.

\emph{Solution of the Abraham-Minkowski controversy of light-atom interacting systems.} In a light-atom interacting system, the interaction between atoms and light is usually treated under electric dipole approximation (EDA), where the coupling between the atoms and the magnetic field of the light is neglected ($\mu=\mu_0$) and the scale of the atom is much less than the wavelength of the light. Therefore, $\mathbf{B}=\mu_0\mathbf{H}$ in such system. The Abraham tenser $\frac{\mathbf{E}\times\mathbf{H}}{c^2}=\frac{\mathbf{E}\times\mathbf{B}}{\mu_0c^2}$ contains no information about the interaction with the atomic degree of freedom, and refers to the Abraham momentum of the light. The other tensor, $\mathbf{D}\times\mathbf{B}$, which contains both the matter and the light degrees of freedom, refers to the Minkowski momentum of light. In fact, the Abraham momentum of light $\mathbf{P_{Abr}}$ in light-atom interacting system becomes $\hbar\mathbf{k_0}/n_g$ \cite{Barnett}; the Minkowski momentum of the light $\mathbf{P_{Min}}$ here is equal to $\hbar\mathbf{k_0}n_p^2/n_g$ for a single photon \cite{PRA}, and eventually equals to $n_p\hbar\mathbf{k_0}$ after the summation of all polariton modes \cite{Barnett}. Since $n_p$ is associated with the real part of the electric susceptibility $Re[\chi]$ by \cite{disper}
\begin{equation}
n_p^2=\varepsilon\mu/(\varepsilon_0\mu_0)=1+4\pi Re[\chi],
\end{equation}
the single-photon's $\mathbf{P_{Min}}$ can be written
\begin{equation}
\mathbf{P_{Min}}=\frac{n_p^2\hbar\mathbf{k_0}}{n_g}=\mathbf{P_{Abr}}+\frac{4\pi\hbar\mathbf{k_0}Re[\chi]}{n_g}.
\end{equation}
Here, the second term is the interaction momentum of the system, which differentiates $\mathbf{P_{Min}}$ and $\mathbf{P_{Abr}}$ of the light \cite{Feigel}, as well as $\mathbf{P_c}$ and $\mathbf{P_k}$ of the atom. This is in good agreement with Eq.(\ref{m}).

The status of an atom can be described by a composite wave function $\psi\varphi$, where $\psi$ and $\varphi$ represent the wave functions of its external and internal degrees of freedom, respectively. For an atom interacting with a coherent light, the wave property of its external motion and internal electric dipole is well conserved in their evolutions, seen from the optical Bloch equations, and the atom will show its $\mathbf{P_c}$ which can be obtained directly from
\begin{equation}
\mathbf{\hat{P}_c}\psi=-i\hbar\nabla\psi=\mathbf{P_c}\psi.
\end{equation}
Conversely, the kinetic momentum operator $\mathbf{\hat{P}_k}$, which includes the interaction momentum term, does not have eigenvalues when acting on $\psi\varphi$, then $\mathbf{P_k}$ is not measurable. Therefore, it has been proved in many works that the system will show $\mathbf{P_c}+\mathbf{P_{Min}}$ with $\mathbf{P_{Min}}=n_p\hbar\mathbf{k_0}$ \cite{OC,OE,EJP,Loudon}. The case for a coherent macroscopic atomic matter wave is similar, since $\psi\varphi$ of each atom has the same phase difference according to the coherent light. The total canonical momentum of the macroscopic matter wave can be obtained as $\mathbf{P_c}=N\mathbf{p_c}$ with $N$ the number of atoms and $\mathbf{p_c}$ the canonical momentum of each atom. A pure Bose-Einstein condensate can be described as a macroscopic matter wave, whose recoil $\mathbf{P_c}$ can be measured precisely due to its ultra-cold temperature. This is the reason why Ketterle's group has observed $\mathbf{P_c}+\mathbf{P_{Min}}$ when BEC is optically pumped and trapped as a coherent macroscopic matter wave \cite{MIT}.

When the external motion of the atoms is spatially incoherent, things become very different. The atoms have space-dependent random phase differences $\phi(x)$ between their external motion and the coherent light (usually laser), which will be multiplied on the composite wave function of the atoms. For $N$ incoherent atoms interacting with a coherent light, with the $ith$ atom having a random phase difference $\phi_i(x)$ to the light, the composite wave function of the $N$ atoms can be written
\begin{equation}
\Psi=\prod_i^Ne^{i\phi_i(x)}\psi_i\varphi_i.\label{sum}
\end{equation}
We can see that $\Psi$ is not the eigenfunction of the canonical momentum operator $\mathbf{\hat{P}_c}=-i\hbar\nabla$ any more due to an additional spatial random term $\sum_i^N\partial\phi_i(x)/\partial x$ appears. Therefore $\mathbf{P_c}$ can not be observed and measured. However, $\phi_i(x)$ is time-independent, which makes the unperturbed Hamiltonian operator $\hat{H}_0$ acting on $\Psi$ still have the eigenvalue as
\begin{equation}
H_0=\sum_i^N\sqrt{\mathbf{p_k}_i^2c^2+{m_i}^2c^4}.\label{fh}
\end{equation}
Here $\mathbf{p_k}_i$ and $m_i$ are the kinetic momentum and proper mass of the $ith$ atom. The external motion of incoherent atoms is closed to its classical limit and the value of $H_0$ is also equal to its classical value, which is the kinetic energy of the atomic cloud. Here one can still obtain the total kinetic momentum of the atoms $\mathbf{P_k}=\sum_i^N\mathbf{p_k}_i$ from the classical value of $H_0$ by Eq.(\ref{fh}), and detect it classically from the center-of-mass motion of the atomic cloud. The system will show $\mathbf{P_k}+\mathbf{P_{Abr}}$ with $\mathbf{P_{Abr}}=\hbar\mathbf{k_0}/n_g$. This is the reason why in many experiments with a laser pulse propagating though hot or room temperature atomic vapor, only the group velocity of light, that corresponding to the $\mathbf{P_k}$ of atoms can be observed.

\begin{figure}
\includegraphics[width=80mm]{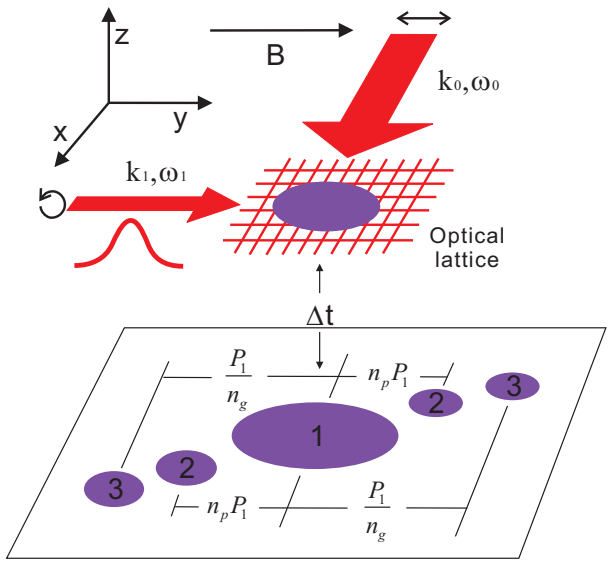}
\caption{(color online) Experimental setup measuring the recoil momentum of ultra-cold $^{87}$Rb atoms by electromagnetically-induced absorption
(EIA). The  strong linearly polarized pump beam ($k_0,\omega_0$) propagates along the $x$ direction and the circularly polarized weak probe beam ($k_1,\omega_1$) propagates along the $y$ direction. At the bottom is the image of atoms' free diffusion delayed by $\Delta t$ after EIA. Region 1 is the pattern of atoms that do not participate in EIA. Region 2 is the pattern of the atoms that obtain a recoil canonical momentum, and region 3 is the pattern of the atoms that obtain a recoil kinetic momentum. $\mathbf{P_1}=\hbar\mathbf{k_1}$ is the free momentum of the probe light, while $n_p$ and $n_g$ are the phase and group refractive index of the atoms, respectively.}\label{config}
\end{figure}

\emph{Experimental setup}. Fig.~\ref{config} shows the experimental setup we designed to test our solution. When the $^{87}$Rb BEC is adiabatically loaded into optical lattices, the system can be described by the Bose-Hubbard model \cite{Bloch}. The ratio between on-site interaction potential $U$ and tunneling potential $J$ determines the ground state of the system. When $U/J\approx0$, the ground state is superfluid state described by a macroscopic wave functions, which is suitable to observe $\mathbf{P_c}+\mathbf{P_{Min}}$ \cite{MIT}. When $U/J>>1$, the ground state is Mott-insulator state, where the atomic number of each lattice site is fixed but the phase of the matter wave on each lattice site is uncertain \cite{Greiner}. Although the short-range phase coherence(among several lattice sites) is still maintained \cite{Mott}, however,the matter wave's long-range phase coherence is collapsed \cite{Bloch}. Since the Mott-insulator state is as cold as BEC and the spatial coherence of the atomic clould is a long range phase coherence, this state is suitable for observing $\mathbf{P_k}+\mathbf{P_{Abr}}$.

\begin{figure}
\includegraphics[width=80mm]{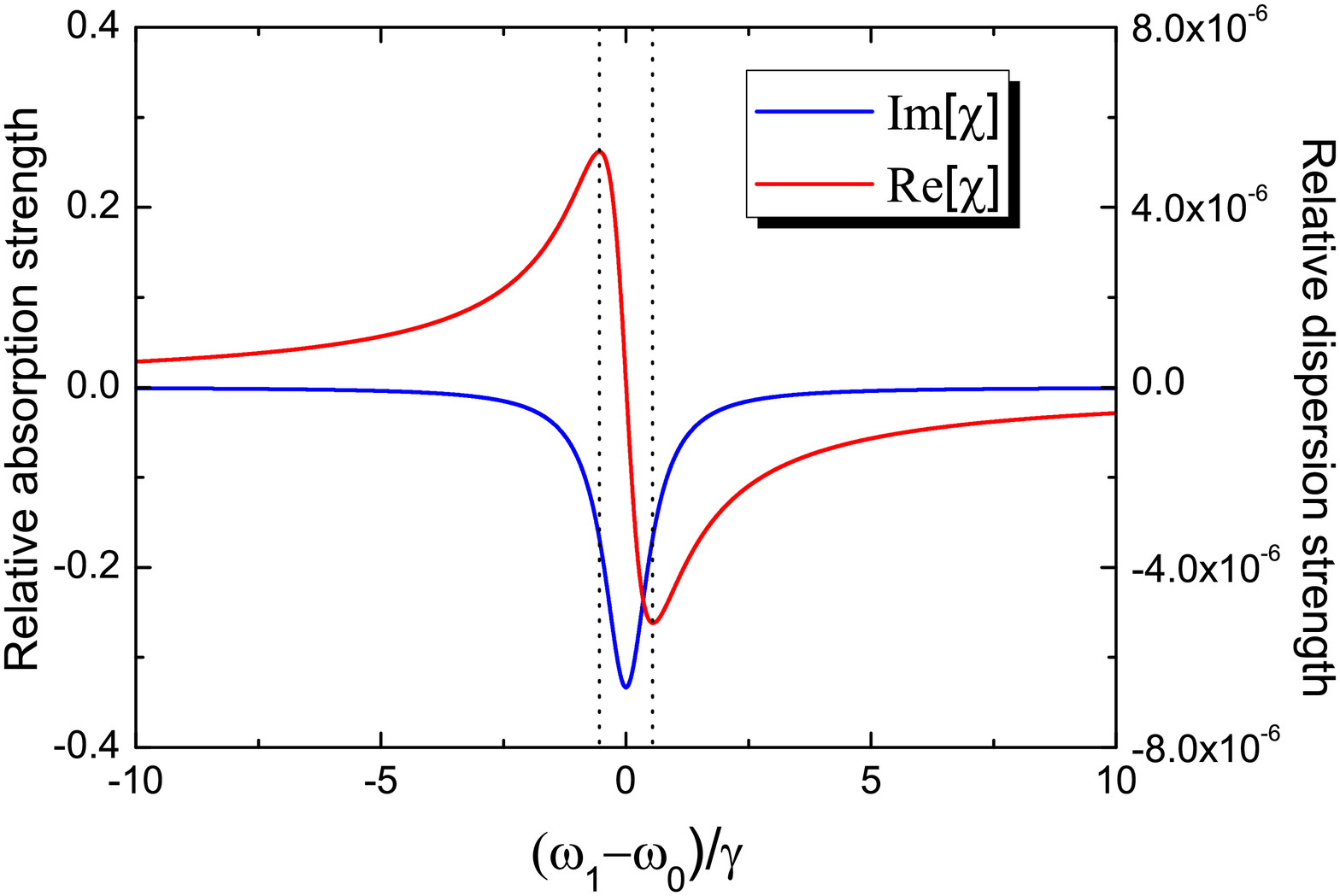}
\caption{(color online) Absorption (blue) and dispersion (red) curves of electromagnetically-induced absorption (EIA) obtained from Kramers-Kronig relation, $Im[\chi]$ and $Re[\chi]$, respectively. $\omega_0$ is the frequency of the strong pump light and $\omega_1$ is the frequency of the weak probe light. The EIA happens when $\omega_1$ is close to $\omega_0$, and the absorption peak is just at $\omega_1=\omega_0$. The width of the anomalous dispersion range is equal to the full-width-at-half-maximum $\gamma$ of the absorption profile.}\label{dis}
\end{figure}

We choose electromagnetically-induced absorption (EIA) \cite{EIA} to make the ultra-cold atoms get recoil momentum from the refracted photons immediately after the optical lattice field is turn off. The atomic density can reach up to $>10^{13}/cm^3$ in red detuning lattices which can provide a good optical thickness. Fig.~\ref{dis} shows the absorption and dispersion curves of EIA, where the EIA signal is a sharp absorption signal whose width is much narrower than the natural width of the $^{87}$Rb's $D_2$ line ($\Gamma=6.056$ MHz). Since $Im[\chi]$ is proportional to the absorption strength and $Re[\chi]$ is proportional to $(n_p^2-1)$, the EIA signal has a strong anomalous dispersion range ($d[n_p(\omega)]/d\omega<0$) due to Kramers-Kronig relation, which can make $d[n_p(\omega)]/d\omega\approx-6\times10^{-11}/$Hz \cite{EIA-ng}.

The ultra-cold $^{87}$Rb atoms are originally trapped in a three-dimensional optical lattice. The pump and probe laser beams for EIA are just set under the optical lattices. The light intensity of the pump beam is around $50 mW/cm^2$, and that of the probe beam is $<0.1 mW/cm^2$. The two beams are vertical to each other to make sure each beam contributes independently to the atomic recoil momentum. The atoms get recoil momentum in $y$ direction only by absorbing refracted photons from the probe beam, and recoiled in $x$ direction only by absorbing refracted photons from the pump beam. The spots of the two beams should be $>0.5 mm^2$ to make the atoms stay for long enough time ($>10 ms$) for the EIA process. The detunings of the two beams should be $5-10$ times larger than $\Gamma$ to maintain the signal strength while avoid D2 line resonant absorption. With an external magnetic field along the $y$ direction, the pump beam is linear polarized in this direction and the probe beam is circular polarized in order to enhance the EIA process, because EIA will be enhanced when the transition type induced by pump and probe beam are not same \cite{EIA-cp}. With this system, we design two sub-experiments to measure the recoil momentum of both spatially coherent and incoherent atoms.

1) The ultra-cold $^{87}$Rb atoms are initially trapped in 3D optical lattices in superfluid state. The pump laser beams ($(\mathbf{k_0},\omega_0)$ are continuous-waves with a narrow width ($<100kHz$). The probe laser $(\mathbf{k_1},\omega_1)$ is a $~10 \mu s$ pulse and is detuned out of the anomalous dispersion range of EIA. At $t=0$, the optical lattices are switched off and the $^{87}$Rb atoms fall into the intersecting region of the pump and probe beams.. The time-sequence should be controlled very well to make the probe pulse just encounter the $^{87}$Rb atoms at this moment. After $\Delta t$, the image of free diffusion will demonstrate that some $^{87}$Rb atoms acquire a recoil momentum $\mathbf{P_c}=n_p(\omega_1)\hbar\mathbf{k_1}$ along the $y$ direction. These atoms will be observed in region $2$ of Fig.~\ref{config}, because the coherent probe light propagates in the coherent matter wave with the Minkowski momentum $n_p(\omega_1)\hbar\mathbf{k_1}$. This sub-experiment functions in the wave-like limit of light-atom interacting system, and the system will present $\mathbf{P_c}+\mathbf{P_{Min}}$.

\begin{figure}
\includegraphics[width=90mm]{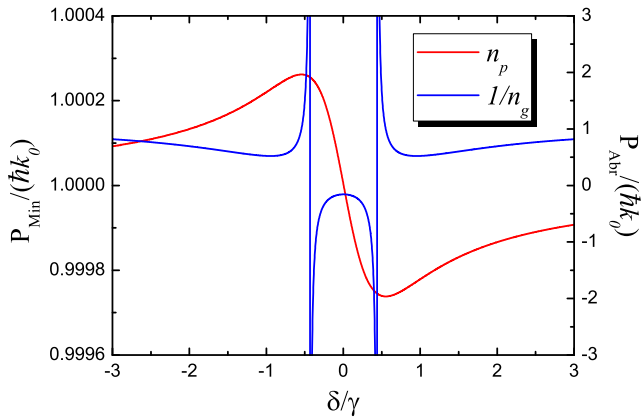}
\caption{(color online) The recoil momentum of cold $^{87}$Rb atoms. The red line is the recoil canonical momentum which is equal to the Minkowski momentum of the probe light, while the blue line is the recoil kinetic momentum that is equal to the Abraham momentum of the probe light. The width of electromagnetically-induced absorption (EIA) signal $\gamma$ is set as 1.0 MHz, and the strength of the signal is set as half of the $^{87}$Rb $F=2\rightarrow F=3$ absorption strength. $\delta=\omega_1-\omega_0$ is the detuning between pump and probe light. Near the two peaks of $n_p$, we can see $1/n_g$ has very large positive or negative values, which may make the distance between region 3 and region 1 much larger than that between region 2 and region 1 in Fig.~\ref{config}}\label{vs}
\end{figure}

2) The ultra-cold $^{87}$Rb atoms are initially trapped in 3D optical lattice in Mott-insulator state.  The pump laser beams ($(\mathbf{k_0},\omega_0)$ are continuous-waves with a narrow width ($<100kHz$). The probe laser $(\mathbf{k_1},\omega_1)$ is a $~10 \mu s$ pulse. At $t=0$, the optical lattices are switched off and the $^{87}$Rb atoms fall into the pump beam. The time-sequence should be controlled very well to make the probe pulse just encounter the $^{87}$Rb atoms at this moment. After $\Delta t$, the image of free diffusion will demonstrate that some $^{87}$Rb atoms get a recoil momentum $\mathbf{P_k}=\hbar\mathbf{k_1}/n_g(\omega_1)$ along the $y$ direction; because the probe pulse propagates in the incoherent ultra-cold atomic ensemble with the Abraham momentum $\hbar\mathbf{k_1}/n_g(\omega_1)$. When the frequency of the probe light is within the anomalous dispersion range, $n_g(\omega_1)$ can be negative due to the dispersive relation
\begin{equation}
n_g(\omega_1)=n_p(\omega_1)+\omega_1\frac{dn_p(\omega_1)}{d\omega_1}<0.
\end{equation}
If $|n_g(\omega_1)|<<1$ (see Fig.~\ref{vs}), a large positive or negative recoil $\mathbf{P_k}$ can be observed with the recoil atoms appearing in region $3$ of Fig.~\ref{config}. This sub-experiment can be considered as the particle-like limit of light-atom interacting system and the system will exhibit $\mathbf{P_k}+\mathbf{P_{Abr}}$.

An advantage of ultra-cold quantum gas is the well controllability of the matter wave coherence. It enables the observation of the interaction between macroscopic matter-wave and light, where both matter and light can be coherent to make $\mathbf{P_c}+\mathbf{P_{Min}}$ possible to be observed. In solid or liquid optical media, things becomes very complicated. Since such condensed matter can not present macroscopic coherence as ultra-cold quantum gas, only $\mathbf{P_k}+\mathbf{P_{Abr}}$ can be measured from their final recoil momenta \cite{ZSU}. However, in the solid or liquid optical media where the electrons are spatially coherent in a wide range, the momentum of light may be close to the Minkowski form \cite{Min_e}. This topic is beyond the content of this article, and we hope quantum simulations with ultra-cold atoms in 3D optical lattices may give some insights to it in the future.

In summary, we find that the Abraham-Minkowski controversy of the light-matter wave interacting system comes from whether the coupling momentum is added on the canonical momentum of the matter wave to form its kinetic momentum, or added on the Abraham momentum of the light to form its Minkowski momentum. The spatial coherence of atoms can determine whether the light-atom interacting system chooses canonical momentum for the atoms and Minkowski momentum for the light, or chooses kinetic momentum for the atoms and Abraham momentum for the light. This is the reason of the Abraham-Minkowski controversy in light-atom interacting systems. Finally, a very realizable experiment is proposed to prove our solution by measuring the recoil momentum of the ultra-cold $^{87}$Rb atoms. Our work enlightens the deep relationship between the momenta and spatial coherence of atoms in the light-atom interacting system. It is useful for future atomic interferometers, cold atom clock, and other precision measurements with cold atoms, where the precise values of the momenta of atoms and light are both required.

This work is supported by NSFC under grants Nos. 10874235, 10934010, 60978019, the NKBRSFC under grants Nos. 2009CB930701, 2010CB922904, and 2011CB921502, and NSFC-RGC under grants Nos. 1386-N-HKU748/10.

\end{document}